# Evolutionary Mechanics: new engineering principles for the emergence of flexibility in a dynamic and uncertain world

James M. Whitacre[1], Philipp Rohlfshagen[1], Axel Bender[2], and Xin Yao[1]
[1] CERCIA, School of Computer Science, University of Birmingham
Birmingham B15 2TT, United Kingdom,
{j.m.whitacre, p.rohlfshagen, x.yao}@cs.bham.ac.uk
[2] Land Operations Division, Defence Science and Technology Organisation
Edinburgh SA 5111, Australia, axel.bender@dsto.defence.gov.au

## Abstract

Engineered systems are designed to deftly operate under predetermined conditions yet are notoriously fragile when unexpected perturbations arise. In contrast, biological systems operate in a highly flexible manner; learn quickly adequate responses to novel conditions, and evolve new routines/traits to remain competitive under persistent environmental change. A recent theory on the origins of biological flexibility has proposed that degeneracy – the existence of multi-functional components with partially overlapping functions – is a primary determinant of the robustness and adaptability found in evolved systems. While degeneracy's contribution to biological flexibility is well documented, there has been little investigation of degeneracy design principles for achieving flexibility in systems engineering. Actually, the conditions that can lead to degeneracy are routinely eliminated in engineering design.

With the planning of transportation vehicle fleets taken as a case study, this paper reports evidence that degeneracy improves robustness and adaptability of a simulated fleet without incurring costs to efficiency. We find degeneracy dramatically increases robustness of a fleet to unpredicted changes in the environment while it also facilitates robustness to anticipated variations. When we allow a fleet's architecture to be adapted in response to environmental change, we find degeneracy can be selectively acquired, leading to faster rates of design adaptation and ultimately to better designs. We also find that the magnitude of fleet design changes are similar between degenerate and decomposable "engineered" options, suggesting that the intelligent introduction of degeneracy needs to be neither costly nor difficult.

In investigating the limitations of degeneracy as a design principle, we consider decision-making difficulties that arise from degeneracy's influence on a system's complexity. While global decision-making becomes more challenging, we also find degeneracy accommodates rapid distributed decision-making leading to (near-optimal) robust system performance. Given the range of conditions where favorable short-term and long-term performance outcomes are observed, we propose that degeneracy design principles fundamentally alter the propensity for adaptation and may be useful within several engineering and planning contexts. In the discussion, we focus on the potential relevance of these findings for the development of new prescriptive guidelines in *complex systems engineering*: a nascent field that applies Darwinian and systems theory principles to improve flexibility and adaptability within systems engineering and planning under uncertainty.

**Running Title:** Evolutionary Mechanics

# 1. Introduction

## Engineering and Planning in Theory

Engineering involves the design and assemblage of elements that work in specific ways to achieve a predictable purpose and function [1] [2]. Engineering, planning, and science in general have historically taken a reductionist approach to problem solving; aiming to decompose a complicated problem into more manageable and well-defined sub-problems that are largely separable or modular, i.e. have an easy to understand (or preferably negligible) inter-dependence with one another (cf Section 3.1 [3], [4]). A reductionist problem decomposition reduces the degrees of freedom that are considered at any one time. It is methodical, conceptually intuitive, and it can help individuals understand the most relevant determinants of each sub-system's behaviour, e.g. by allowing one to readily discover confounding factors and eliminate spurious associations. If sub-systems truly represent modular building blocks of the larger system, then there are (by definition) relatively few ways in which the sub-system will interact with its surroundings. This often permits the operating environment to be defined with precision and accuracy. When these conditions are met, engineers and planners have historically been able to systematically design components/subsystems with reliable functioning that translates into reliable performance at the system-level [5]. The application of reductionist principles results in a hierarchical decomposition of a system that contributes to system-level transparency and benefits global forms of decision-making, trouble-shooting, planning, and control.

## Engineering and Planning in Practice

While the reductionist paradigm is logically sound, it is only loosely followed in practice because many of the systems within which engineering (or reengineering) takes place cannot be neatly decomposed. This is at least partly due to the factors that have shaped these systems over time including bounded rationality, distributed (partially decoupled and concurrent) decision-making, multiple conflicting objectives, historical contingency (path dependency), and environmental volatility.

While engineers have been aptly described as "tinkerers" in this context [6] [7], purest views of planning and engineering often prevail when attempting to understand failure in complex systems. When failure occurs, it is common and logical to highlight the precise points where a system has failed with the implicit assumption that performance could have been sustained had a relatively small set of modifications been made during its design. Because a narrative description of failure can be achieved through careful investigation, it is often assumed that the precise contingency of failure sufficiently captures the origins of a system's design flaws. That these failures might be symptomatic of wider issues related to innate system properties is sometimes suggested but has been notoriously difficult to address in practice [8] [9] [10].

A further limitation to the classic reductionist paradigm is that it assumes the capacity to anticipate beforehand the conditions a system will experience and the precise manner in which the system should respond to those conditions [3], i.e. it assumes a stable environment or it requires precognition [11]. While predicting plausible future conditions can often be a useful exercise, several factors can limit prediction accuracy and lead to uncertainty [5]. The origins of this uncertainty are varied, however it is a general rule of thumb that for complex systems operating in a dynamic environment (e.g. socio-technical-ecological-economic systems, enterprises, systems-of-systems), we are limited in our ability to develop standard operating procedures or contingency plans that can accurately account for the various plausible conditions that we might encounter [12] [13] [14]. When conditions cannot be fully anticipated in advance, it becomes important that a system be adaptable to novel conditions that it was not explicitly designed to address. From a classic engineering and planning perspective where components are designed for a pre-specified purpose, such a stated design goal is ambiguous and vague and it may seem that planning for the unexpected is an oxymoron.

There is a rich history of (non-reductionist) systems research in engineering and organization science dating back to the work of Jay Forrester at General Electric [15]. More recently a number of systems engineering and organization science studies have proposed ways to exploit properties commonly associated with biological robustness and adaptability [16] [21] [22] [23]. While a better appreciation of these biological properties has been useful to decision makers and planners, many of these properties - including loose coupling, distributed robustness, and adaptability - are hard to apply because there is little understanding of their origins or mechanistic basis and few guidelines for their realization. One of the key aims of this work is to propose and validate design principles that can be defined at a component level and that can lead to the emergence of system properties such as flexibility, distributed robustness and adaptability. Because these design principles correspond with measurable properties that can be realized in an engineering context, in this article we are able to validate our most important claims through simulation case studies involving engineered systems that undergo design optimization using evolutionary algorithms.

# Lessons from Biology

The creation and maintenance of biological functions (e.g. traits, services) occurs in a non-reductionist manner that is exceptionally effective at accommodating and exploiting novel conditions. By distilling out basic working principles that lead to biological robustness and adaptability, we hope to provide useful insights or even design principles that can be used within engineering and planning contexts. Here we focus on one particular design principle that is suspected to play a fundamental role in assisting complex biological systems to cope with environmental novelty. In introducing this topic, we limit the discussion to essential concepts and refer the interested reader to the more thorough discussions in [24] [25] [26]. Given the multi-disciplinary nature of our topic, this introduction relies on abstract terminology that can be universally understood by biologists, engineers, planners, and decision-makers. Later on in the Case Study and Discussion sections (Sections 2 and 5), we establish links between these abstract concepts and their counterparts within particular disciplines.

### Robustness: origins

Engineered systems typically comprise elements designed for a single and well-specified purpose, however in biological systems there is no pre-assignment of a one-to-one mapping between elements and traits. Instead, at almost every scale in biology, structurally distinct elements (e.g. genes, proteins, complexes, pathways, cells) can be found that are interchangeable with one another or are otherwise compensatory in their contributions to system functions [27] [25] [24] [26] [28] [3] [29] [30] [31] [32] [33] [34]. This many-to-one mapping between components and functions is referred to as *functional redundancy*.

Pure redundancy is a special case of many-to-one mapping and it is a commonly utilized design tool for improving the robustness of engineered systems. In particular, if there are many copies of an element that perform a particular service then the loss of one element can be compensated for by others; as can variations in the demands for that service. However, maintaining diversity amongst functionally similar elements can lead to additional types of stability. If elements are somewhat different but partially overlap in the functions they perform, they are likely to exhibit different vulnerabilities: a perturbation or attack on the system is less likely to present a threat to all elements at once. Alternatively, we might say that a system gains versatility in how a function can be performed because functionally redundant elements enhance the diversity of conditions under which a particular demand can be satisfied [24] [35] [36].

### Functional plasticity and buffering efficiency

Whether elements are identical (i.e. *purely redundant*) or only functionally redundant, the buffering just described always requires an excess of resources, and this is often viewed in engineering as a necessary but costly source of inefficiency [7] [37] [9]. What is less appreciated however is that simple trade-offs between buffering and efficiency do not necessarily arise in biological systems. Not only are different components able to perform the same function (a many-to-one mapping), many of these components are also multi-functional (one-to-many mapping), with the function performed depending on the context; a behavior known as *functional plasticity* [25] [38] [28] [39] [40] [41] [42]. Functionally plastic elements that are excluded from participating in a particular function (e.g. due to demands for that service already being met) will switch to other functions where they can be utilized instead [28] [43]. Functional plasticity thus alters the tradeoff between efficiency and buffering because excess resources are shared across multiple tasks.

If functionally plastic elements share similarities in only some of their functional capabilities, then the timing and nature of functional demands placed on these components will not be identical. A difference in the timing of demands leads to differences in availability, which can increase the variety of conditions that a system can perform a particular function. This can alter a system's robustness because stability depends on the ability to elicit a particular response or achieve a particular function at the time and place where it is needed.

Functional plasticity and functional redundancy are observed at all scales in biology and we have described simple and generic situations where these properties contribute to trait stability through local compensatory effects. In biology, degeneracy -partially overlapping functionality amongst multi-functional components- is a more commonly used term for describing the co-occurrence of functional redundancy and functional plasticity. In the literature, there is an extensive list of documented cases where degeneracy has been found to promote trait stability [27] [24] [26] [25].

More importantly however, degeneracy has also been shown to create emergent system properties that further enhance trait stability through *distributed compensatory effects*. The origins of this additional and less intuitive form of robustness have been described in the *networked buffering hypothesis* [24]. Using genome:proteome simulations, Whitacre and Bender found evidence that networked buffering can roughly double the overall robustness potential of a system for each of the perturbation classes tested [26] [24]. An important conclusion from those studies is that even small amounts of excess functional resources can have a multiplicative effect on system-level flexibility and robustness when degeneracy is prevalent in a system [26] [24]. This is of particular relevance to this study because the emergence of these distributed forms of robustness has also been found to substantially alter a system's ability to adapt to novel conditions. Before discussing how degeneracy influences a system's adaptation capabilities, we first review important requirements for the discovery of new adaptive improvements in a system.

### Adaptability: The disposition to adapt
**Accessing novelty**: In both biology and engineering, the discovery of an improved component design necessitates the exploration of new design variants. Theoretically, degeneracy should enhance a system's access to design novelty because functionally redundant elements retain unique structural characteristics. Structural differences afford a multiplicity of design change options that can be tested, and thus provide more opportunities for new innovative designs to be discovered [25] [26] [44] [45] [46] [47].

**Transforming novelty into innovation**: The availability of distinct design options is an important prerequisite for innovation, however new opportunities often come with new challenges. To transform a local novelty into an exploited innovation, a system must be flexible (e.g. structurally, behaviourally) to accommodate and utilize a modified component effectively. For instance, design changes sometimes require new specifications for interaction, communication, operating conditions, etc. However, a system must accommodate these new requirements without losing other important capabilities or sacrificing the performance of other core system processes. In other words, the propensity to innovate is enhanced in systems that are robust in their core functions yet flexible in how those functions are carried out [45].

**Facilitating unexpected opportunities**: Because design novelty is not predictable, the flexibility needed to exploit design novelty also cannot be pre-specified based on the anticipation of future design changes. To support innovation, it appears that this robust yet flexible behaviour would need to be a property that is pervasive throughout the system. Yet within the wider context of a system's development – where each incremental design change involves a boundedly rational and ultimately myopic decision – it also seems that this flexibility must be an emergent property that, while intimately tied to the history of development, can readily emerge without foresight.

We have established a theoretical basis explaining how degeneracy can support these prerequisites for adaptation and we have recently accumulated some evidence from simulation studies that supports these conjectures. For instance, we have found evidence that degeneracy considerably enhances access to design novelty (see heritable phenotypic variation in [26] [48]). We have also shown that these novelties can be utilized as positive adaptations [49] [50] and can sometimes afford further opportunities when presented with new environments [51]. In attempting to understand how novelties are transformed into adaptations, we have shown in [24] that high levels of degeneracy lead to the emergence of pervasive flexibility in how a system can organize its resources and thus allows a decoupling between the preservation of some functions and the accommodation of new ones. These experimental findings support each of our stated conjectures regarding how degeneracy provides a mechanistic basis for robustness and adaptability in biological systems [27] [26] [25].

### Conflicts between robustness and adaptability
The requisite conditions that we have outlined above suggest a relationship between robustness and adaptability that we argue is rarely observed in engineered systems and may even be entirely absent in systems that are designed entirely using classic reductionist design principles. Some evidence supporting this view is found by comparing the relationships between robustness and adaptability in human-designed evolutionary optimization algorithms with that observed in natural evolution. In agreement with theories on neutral evolution [52] [53] [54], simulations of gene regulatory networks and models of other biological systems have indicated that increasing mutational robustness may increase a system's propensity to adapt [46] [26]. Taking cues from the original theories, research into nature-inspired optimization has looked at designing mutational robustness into an optimization problem's representation and has almost exclusively done so through the introduction of gene redundancy, e.g. polyploidy. The result has been a negligible or sometimes negative influence on the adaptive capabilities of individuals and populations [55] [56] [57] [58] [59] [60] [61]. More recently we have investigated simulations where evolution is given the option to create robustness through degeneracy. We have found evidence that this leads to the selective growth of degeneracy, substantial improvements in evolving robustness towards environmental volatility, and better adaptive capabilities in responding to environmental novelty [49] [50]. Thus, positive relationships between robustness and adaptability have historically been absent in evolutionary simulations but can be established when designed redundancy is replaced by the evolution of degeneracy.

### Summary and Outline

The previous sections reviewed a set of properties that differ between purely evolved (bottom-up) systems and purely designed (top-down) systems and we have proposed that these properties help to explain how biological systems cope with novel conditions. The theories that underpin these ideas are conceptually straightforward [27] [25] yet also operationally useful as they position degeneracy as a mechanistic facilitator of robustness and adaptability that, in principle, could be applied outside biological contexts [25].

As a first step towards investigating the merits and limitations of degeneracy as an engineering design concept, in the following sections we introduce a strategic planning problem that meets each of the requisite conditions for degeneracy to emerge and where the ability to cope with novelty is intimately tied to success. To explore degeneracy design principles, we modify simulation conditions from [49] in order to more closely reflect systems engineering problems subjected to heterogeneous and unpredictable environments. While we previously reported some evidence of positive associations between degeneracy, robustness, and adaptability in [49] [50], our aim in the current study is to determine under what conditions degeneracy will confer advantages to systems that are responding to novel conditions. Here we report certain classes of environmental dynamics as well as cross-scale systemic interactions that influence performance gains from degeneracy and are suggestive of intuitive (and theoretically anticipated) bounds on the utility of degeneracy design principles. In discussing these results, we comment on other aspects of our simulations that should be explored to further clarify the conditions under which degeneracy might be useful as a design principle. Our findings suggest that degeneracy provides advantages under experimental conditions that overlap with several systems engineering contexts.

Following a brief introduction of long-term planning problems, we present a strategic planning problem involving investment decisions for a fleet of vehicles that face considerable uncertainty in their future operating environments. We introduce a simulation environment for evaluating fleet robustness across future plausible environments (scenarios) and an evolutionary algorithm that is used to simulate incremental adaptations that are aimed at improving the robustness of a fleet's design. In Section 4, we explore several robustness and adaptability properties of the fleets as they are exposed to different classes of environmental change and we find evidence supporting the hypothesis that degeneracy provides broad advantages within some classes of environments; most notably environments that are complex and occasionally unpredictable. We explore some of the limiting conditions that determine when benefits are derived from degeneracy, focusing particularly on classes of environmental dynamics (decomposable/predictable vs. uncertain changes) and system dynamics (local vs. global decision-making). Section 5 comments on the potential relevance of these findings and we give concluding remarks in Section 6.

## 2. Case Study

### Strategic Planning Overview

In strategic planning problems, uncertainty arises from the long time horizons in which planning goals are defined and the properties of the systems and environments in which planning takes place. In particular, strategic planning almost invariably deals with the manipulation of multi-scaled complex systems, i.e. systems comprising many heterogeneous elements whose actions and interactions translate into emergent system functions or *capabilities* that are observable over several distinct timescales. A number of well-known contemporary issues are exemplars of strategic planning problems and include financial market regulation, strategies for responding to climate change and natural disasters, development assistance to third-world countries, defence planning, and strategic planning for nations and large organizations.

Uncertainty within strategic planning can be characterized in different ways [62] [63] [64] [65]; hence many differing uncertainty taxonomies exist (see, e.g., [66]). For instance, uncertainty can be classified into "known unknowns" (conscious ignorance), "unknown knowns" (tacit knowledge) and "unknown unknowns" (meta ignorance) [67]. Different classes of uncertainty pose different challenges, and traditional approaches to planning have mostly been developed to address conscious ignorance, e.g. through prediction modelling, process control etc. Tacit knowledge and meta-ignorance cannot be explicitly planned for and require the development of system properties that facilitate adaptation at different levels within a system. For instance, tacit knowledge is predominantly revealed during plan execution and exploiting this knowledge requires plans to be responsive (behaviourally adaptive). On the other hand, meta-ignorance often gives rise to shocks or surprises and knowledge about the "unknown unknowns" only reveals itself "after the fact". Dealing with meta-ignorance therefore requires adaptable design and behaviour that allow for innovation to germinate. Past experiences only provide very limited useful insights.

The existence of meta-ignorance means that we are never entirely certain what future events might be encountered or how current decisions will shape future events or ourselves [68]. When we don't know what we don't know, we are prevented from formulating our uncertainty based on the likelihood of possible future states; a common assumption used in robust optimization and control. Even the articulation of plausible states can be difficult due to the emergence of new phenomena. While many aspects of our world remain conserved over time (described in the literature as historical contingency, path dependency, or "frozen accidents" [69] [70] [71] [10], but also see [8] [45]), other aspects are not predictable due to, for instance, unanticipated regime shifts in the physical environment, paradigm shifts in knowledge and culture, and disruptive technological innovations that introduce new opportunities as well as new challenges. Under these circumstances, desirable solutions or strategies should not only be robust to expected variability; they should also have an innate propensity to adapt to novel conditions. More generally, it is important to understand what design principles, routines, and system properties will determine a system's capacity to function under high variability (robustness) and facilitate adaptations to unexpected novelty, e.g. in the environment. Because many long-term planning problems are appropriately modeled using semi-autonomous and multi-functional agents (each of which are also complex systems), this class of problems provides a suitable domain for exploring the biologically inspired principles we presented earlier.

The next section introduces a strategic planning problem involving investment decisions in a fleet of military field vehicles. Each of the general issues that we have raised here are also directly relevant for this particular problem. For instance, over short (operational) timescales, fleets must operate within a volatile environment; having the capacity to rapidly deal with both anticipated and unanticipated missions. Over longer timescales, environments change more dramatically and it is important that new fleet architectures can be developed that can adequately respond to these new realities.

## 3. Experimental Setup

### Model Overview

In this paper we consider a simple but well-defined framework that allows us to fully study the system properties in which we are interested. The most important dynamics captured by these simulations include changes in the assignment of vehicles to tasks, changes in the composition of vehicles that make up a fleet, changes in the tasks that must be executed during fleet operations, and changes in the composition of tasks at a timescale comparable to the timescale for changes in fleet composition.

The problem is specified by a *land fleet model* and is based on earlier work carried out by Whitacre and Bender in [48]. The model consists of a *fleet* of vehicles and each vehicle corresponds to a specific *vehicle type* that is capable of carrying out *two* specific *task types*. Each vehicle may devote its resources (e.g., time) to either or both tasks: a vehicle that is capable of performing tasks *A* and *B*, for instance, could devote 30% and 70% of its time to task *A* and *B* respectively. The fleet is exposed to a number of environmental conditions (referred to as *scenarios*), each of which specifies a set of demands for the specific task types. The fleet responds to each environment by re-distributing its resources using a local decision making process. In particular, each vehicle may distribute its own resources amongst its two tasks, based on global feedback about the current priority level for fulfilling each task. The problem can thus be stated as finding a suitable fleet composition (i.e., set of vehicle types) that allows the fleet to respond to a volatile environment through the redistribution of its resources. In our experiments, this fleet composition (or architecture) is evolved, using a genetic algorithm, to maximize fleet robustness towards this environmental volatility. In order to compare and evaluate different fleet architectures, constraints are imposed on the evolutionary process so that degeneracy is allowed to emerge in some but not all cases.

# Simulation Details

### Vehicle fleet & task types
The problem is formally defined as follows: There are a total of *m* task types and vehicles are constrained in what types of tasks they can perform. This constraint emulates restrictions that may arise naturally in real-world scenarios where specific tasks have specific hardware requirements that may either be in conflict with one another or simply too expensive to be combined within a single vehicle. In particular, we constrain task-type combinations by defining a symmetric neighbourhood for any task type $0 < i \leq m$ as $N_r(i) = \{j \mid |i - j| \leq r, i \neq j\}$ where $r$ is a predefined *radius*, set as $r=2$ in all experiments. We assume the set of task types to have periodic boundaries such that the neighbourhood wraps around the extreme values.

The design of a fleet of $n$ vehicles is characterised by the matrix $V = \{0,1\}^{n \times m}$: If vehicle $V_i$ can carry out task-type $j$, then $V_{ij} = 1$. Otherwise, the matrix entry $V_{ij}$ is 0. As any vehicle is capable of exactly two types of tasks, it follows that $\sum_{j=1}^{m} v_{ij} = 2, \forall i$. Furthermore, task combinations in any vehicle $i$ are constrained such that for any $j$ and $k$ for which $V_{ij} = V_{ik} = 1$, it follows that $k \in N_r(j)$. The matrix $V$ thus fully specifies a valid fleet architecture where each vehicle is constrained in (a) the number of task types it may execute and (b) the relationship between the task types (i.e., the radius).

A second matrix $R = \{x \in \mathbb{Z} : 0 \leq x \leq 10\}^{n \times m}$ is used to indicate the degree to which each vehicle distributes its resources. Each element $R_{ij}$ specifies how much of its resources vehicle $V_i$ has assigned to task type $j$. Any entry $R_{ij}$ may be non-zero only if vehicle $V_i$ is able to perform task $j$ (i.e, $V_{ij} = 1$). We assume that a vehicle allocates all of its resources such that $\sum_{j=1}^{m} R_{ij} = 10, \forall i$. The alteration of elements in $R$ thus specifies the distribution of tasks across the fleet of vehicles.

### Redundancy & degeneracy
The matrix $V$ specifies the task types each vehicle in the fleet can carry out. We may thus use $V$ to compute a degree of redundancy or degeneracy found in the fleet. Two vehicles are redundant with respect to each other if they are able to carry out precisely the same two task types. If two vehicles have exactly one task type in common, they are considered degenerate. In order to calculate redundancy and degeneracy in a fleet, we consider all $(n^2 - n)/2$ unique pair-wise comparisons of vehicles and count the number of vehicle pairs that have identical task capabilities, partially similar task capabilities and unique capabilities. We then normalize these measurements by the total number of pair-wise comparisons made. More specifically, vehicles $V_i$ and $V_j$ are redundant if $V_i \cdot V_j = 2$, degenerate if $V_i \cdot V_j = 1$ and unique with respect to one another if $V_i \cdot V_j = 0$. The fleet's degree of degeneracy is then given by:

$$|V|_{degen} = \frac{1}{n^2 - n} \sum_{i=1}^{n} \sum_{\substack{j=1 \\ j \neq i}}^{n} [V_i \cdot V_j = 1]$$

where [·] returns 1 if the containing statement is true and 0 otherwise.

### Environments
At any moment in time, $t$, the fleet is exposed to a set of $\varepsilon = 10$ environmental scenarios $E_t = \{e_1^t, e_2^t, \ldots, e_\varepsilon^t\}$ where $0 \leq e_{ij}^t \leq n$ specifies the number of times task type $j$ needs to be executed in scenario $e_i$ at time $t$. For each scenario $e_i^t$, the complete satisfaction of task requirements requires a fully utilized fleet[1], i.e. $\sum_{j=1}^{m} e_{ij}^t = 10n, \forall i, t$. The scenarios are generated as follows: The first scenario in the set — the "seed scenario", $e_{seed}$ — is created by randomly sampling, with replacement, $n$ tasks from the $m$ task types. The remaining $\varepsilon - 1$ scenarios are generated from the seed using *two* methods as described below. We distinguish between a decomposable case, where correlations are restricted to occur between predetermined pairs of tasks, making the problem separable, and a non-decomposable case where correlations between the variation of any two task types is avoided.

### Decomposable set of scenarios
In order to generate the decomposable scenarios, the $m$ task types are randomly grouped into $m/2$ pairings which are place into a set $S$. A random walk is then used to create new scenarios based on $e_{seed}$. At each step of the random walk, a single task $i$ is selected and its demand incremented by 1. At the same time, the demand for the corresponding task $j$, i.e. the task type for which $(i, j) \in S$, is decremented by 1, ensuring that the total number of tasks required by the environment is always constant and equal to $10n$. The correlated increase/decrease is allowed as long as either task demand is within the bounds of 0 and $n$. The *volatility* of $E_t$ is subsequently defined as the length of the random walk which always starts from $e_{seed}$.

---
[1] Assuming the architecture of the fleet is optimal with respect to the environment (i.e, the fleet *can* satisfy all demands given an optimal distribution of resources).

**Non-decomposable set of scenarios**
In the non-decomposable case, each step of the random walk involves an entirely new and randomly selected pair of task types (i.e., there is no predetermined set $S$). This ensures that the number of correlated task types is minimised, making the problem non-decomposable.

**Validation scenarios**
For selected experiments, fleet robustness evolves for one set of environmental scenarios ("training scenarios") and is then further evaluated on a new set of "validation scenarios". The validation sets are generated in the same manner as the scenarios used during evolution but with the following additional consideration: In each validation set, one scenario is selected at random from the set used during evolution and is then used as the seed to generate the new set of scenarios. For validation set 1 ("Validation I"), the generation of the remaining scenarios is further constrained such that variations between the original and validation set remain decomposable, i.e. the same task type correlations (as of set $S$) are used. For validation set 2 ("Validation II"), the remaining scenarios are generated without such restriction, i.e. they are non-decomposable. This implies that the scenarios within the second validation set are not only non-decomposable with respect to one another but they are also non-decomposable with respect to the training scenarios.

**Evaluating fitness**
The fleet of vehicles is specified by the matrix $V$ and the distribution of resources is given by the matrix $R$. The fitness of a fleet $V$ at time $t$ is determined by how well the fleet satisfies the demands imposed by the set of environments $E_t$. For every scenario $e_i^t \in E_t$, the fleet may compute a new matrix $R_i^t$. We denote as $\delta_j(R_i^t) = \sum_{k=1}^{n} R_{i\ kj}^t$ the sum of elements across column j in matrix $R_i^t$, which corresponds to the total amount of resources the fleet devotes to task type $j$. The fitness of the fleet with respect to environment $e_i^t$ is then calculated as follows:

$$f_{e_i^t} = \sum_{j=1}^{m} max\{0, e_{ij}^t - \delta_j(R_i^t)\}^2$$

Fleets are exposed to $\varepsilon$ scenarios at any one moment in time and the fitness of a fleet is simply the average over the individual fitness values:

$$F_{E_t} = \frac{1}{\varepsilon} \sum_{i=1}^{\varepsilon} f_{e_i^t}$$

It should be noted that some fleet $A$ is considered fitter than some fleet $B$ at time $t$ when $F_t(A) < F_t(B)$.

The fitness of a fleet depends on its composition and the subsequent construction of $R$. The resources are allocated using a local asynchronous decision making process that aims to optimize fitness: all vehicles are considered in turn and for each vehicle, its resource allocation is adjusted (using ordered asynchronous updating). We increment/decrement the value $R_{ij}$ according to how this affects the fleet's fitness (global feedback). Once no further improvements are possible, the next vehicle is considered. This process is repeated until no further improvements can be made.

# Genetic Algorithm
To evaluate the merits and limitations of reductionist versus degenerate design principles, we define specific design options for a fleet such that certain structural properties can be maintained within the fleet architecture. The resource allocation of the fleet is computed on the fly (as tasks are introduced) but the fleet's composition (i.e., its vehicle types) needs to be optimized towards a particular $E_t$. In order to do so, we employ a genetic algorithm based on deterministic crowding. The fleet is represented as a vector of vehicle types. During the evolutionary process, two parents are randomly selected from a population of 30 fleets (without replacement) and subjected to uniform crossover with probability 1: The crossover operator selects, with probability 0.5, for each locus along the chromosome, whether the corresponding gene (vehicle type) will be allocated to offspring 1 or 2. Each offspring is mutated (see below) and then replaces the genotypically more similar parent if its fitness is at least as good.

The design options available to a fleet (i.e., redundancy or degeneracy) are controlled exclusively by the mutation operator: When reductionist design principles are enforced by the mutation operator, the fleet always maintains a fully decomposable architecture. This is illustrated in Figure 1 where fleets comprise vehicles that are either identical (purely redundant) or entirely dissimilar in their functionality. Unsurprisingly, fleets with this architecture acquire much of their robustness through vehicle redundancy and consequently are referred to as *redundant fleets*. When reductionism is not enforced by the mutation operator, some changes to vehicle designs may result in a partial overlap between the capabilities of vehicles (see Figure 1). Fleets evolved under these conditions are referred to as *degenerate fleets*. In our experiments, the initial fleet is always fully redundant independent of the mutation operator used.

The mutation operator plays a central role as it is used to restrict the fleet compositions that may evolve. Its design is thus crucial, especially as it is important to ensure that both fleet compositions, redundant and degenerate, are obtained due to selective pressure alone and not auxiliary effects such as differences in solution space size. The mutation operator has thus been designed with the following considerations in mind: (a) the search space is to be of the same size in all experiments; (b) in some experiments both redundancy and degeneracy can be selected for during the evolutionary process (as opposed to degeneracy emerging as a consequence of the specifications of the model).

The mutation operator replaces exactly one randomly chosen vehicle in the fleet with a new vehicle type. For each locus in the chromosome, replacement options are predetermined and limited to $m/2$ unique vehicle types. For experiments in which the fleet cannot evolve degenerate architectures, alleles for all loci are drawn from set *S*. It follows that a purely redundant fleet remains redundant after mutation. For experiments in which the fleet can evolve degenerate architectures, allele options are almost the same as before, except that for half the alleles available at each locus, one task is changed to a new task type, thus allowing a partial overlap in functions to be possible between genes in the chromosome.

### Construction of un-evolved degenerate and redundant fleets

In some of our experiments we compare the results obtained from evolving fleets (using the genetic algorithm) with "un-evolved" fleets that have the same composition characteristics as the evolved fleets. In constructing an un-evolved fleet, we proceed with an initial fleet as defined at the beginning of our evolutionary simulations, i.e., a randomly generated redundant fleet. We then proceed to iteratively implement the mutation operators associated with each fleet class (redundant, degenerate) with the mutated fleet replacing the original if its redundancy or degeneracy measurement becomes more similar to that of the evolved fleets. This mutation/selection procedure is iterated until redundancy or degeneracy differences are less than 5%.

## Robustness

The term robustness bears different meanings depending on the context in which it is applied, however most definitions relate to the insensitivity of a functional output or system trait when a system is exposed to a set of distinct conditions. While the set of *distinct conditions* are problem-specific, most involve small internal changes (e.g. in vehicle design parameters) or small changes in the external environment (e.g. changes in the conditions in which a fleet's fitness is evaluated). For many domains, including the current case study, robustness is intimately tied to system fitness and is characterized as the maintenance of good fitness values, e.g. an aggregation function of fitness values $F_{E_t}$ (Eq. 3) with $F_{E_t} < T$ where $T$ describes some satisfactory fitness threshold.

One drawback to such measurements arises when the threshold $T$ is high. Then complete and immediate failure is unlikely. While many fleets will "survive" or appear robust to most conditions, they may still exhibit substantial differences in their ability to maintain competitive performance with important subsequent effects to their long-term survival or growth prospects, i.e. there is a decoupling between robustness over short and long timescales. On the other hand, with tolerance thresholds $T$ too low, all fleets will fail to meet the fitness requirements and will appear equally poor. This poses a technical challenge to fleet optimization as it eliminates selection pressure if corrective steps are not taken, e.g. thresholds are continually changed over time to ensure distinctions are possible in comparing fleets that would otherwise be consistently below or above satisfactory levels. One alternative is to create a two-tiered selection criterion whereby fleets are first compared based on satisfying fitness thresholds, and when differences are not resolved by this comparison they are further compared by their average fitness over all scenarios. This is the approach that we will take in the analysis of our experiments. When using this measurement in preliminary tests we found it was still necessary to tune the threshold (or tune scenario volatility) in order to resolve differences between fleets.

While our findings might change when scenarios are assigned different priorities or when the number of scenarios is small (< 20), our tests indicate that reporting an average fitness generates qualitatively similar conclusions in comparison to the use of a tuned robustness measurement. Average fitness is not a commonly used measure of robustness, however it does provide some advantages in the current study. In particular, an average fitness allows for a direct observation of abrupt performance changes in a system that could otherwise be dismissed as minor differences attributable to the crossing of a threshold.

## 4. Results

### Evolution in Environments with Decomposable Volatility

In the first set of experiments, we evolve fleets of vehicles to effectively handle a set of anticipated future environments or scenarios. Each scenario defines a set of tasks that the fleet must perform and scenarios differ from one another in the frequency of task requirements. We proceed by investigating problems where environmental uncertainty is highly constrained and where reductionism in fleet design is expected to be favoured. In particular, fleets are evolved in environments where, unbeknownst to the fleet optimization algorithm, the differences between scenarios exactly match vehicle capabilities within a fleet of a specific structural decomposition (i.e. a particular redundant fleet). We call these "scenarios with decomposable variation".

With environmental variations matching a decomposable architecture, any fleet with that same decomposition will find that these variations can be decomposed into a set of smaller independent sub-problems that each can be addressed by a single class of vehicle and thereby solved in isolation from other vehicle types. This also means that any changes in fleet design that improve performance for one scenario will not result in lowered performance in other scenarios, i.e. performance across scenarios is correlated and evolution proceeds within a unimodal fitness landscape. These features provide ideal conditions for reductionist engineering design and we see in Figure 2a that such design approaches can generate maximally robust fleets. Notice that according to our definition of fleet fitness, $F_t$, in Eq. 3 a fleet is considered more robust than another fleet when its robustness measure is *smaller*. "Maximum" robustness is achieved when our robustness measure approaches 0.

Allowing fleet architectures to deviate from a pure decomposition should not be beneficial for these types of predictable problems. While the environment varies in an entirely decomposable manner, available fleet responses do not. During fleet evolution, we see in Figure 2a that fleets with degeneracy initially evolve quickly and can learn how to respond to most variations within the environment, although evolution does not discover a maximally robust architecture in every problem instance (see "Training" results in Figure 2b).

### Robustness in Environments with Uncertainty

We next evaluate how the evolved fleets respond to various forms of environmental uncertainty. In particular, fleets evolved under the previous set of environments (now called *training scenarios*) are reevaluated in a new set of environments (*validation scenarios*). Uncertainty is characterized by the differences between training scenarios used during evolution and the new scenarios that are defined within the validation set. We present results for two validation tests that allow us to highlight some of the strengths and weaknesses that we have observed in the two fleet architectures.

**Validation I:** As described in Section 3, the first validation set is generated by selecting a scenario from the training set and using its position in scenario space as a seed for generating a new set of scenarios, i.e. with all subsequent validation scenarios generated by taking short random walks away from the seed position in scenario space. The intuition behind this validation test is that actual future conditions are often partially captured by our expectations such that some expectations/scenarios are invalidated while other regions of scenario space become more relevant and are expanded upon. We impose the following additional constraints on validation set I: differences between scenarios must be decomposable when comparing scenarios within the validation set as well as across the training and validation sets.

While redundant fleets are not guaranteed to optimally respond to every new scenario, the constraints imposed on the validation set should place these fleets at a considerable advantage. It is seen in Figure 2b that redundant fleets often maintain optimal robustness however they also occasionally exhibit poor performance outcomes. While the parameters used to generate the validation scenarios will influence the ratio between poor performance outliers and optimal robustness, we found bimodal validation responses were a common phenomena in redundant fleets. In particular, redundant fleets tended to provide highly effective responses towards scenarios that fit within their design limits but once these limits were exceeded, their performance degraded markedly.

In contrast, the degenerate fleets are not designed in a manner that is biased towards the validation set's decomposition. In many validation instances the fleet's robustness is degraded as its repertoire of responses fails to match the new requirements (Figure 2b). However, performance losses were relatively attenuated, i.e. degenerate fleets were less sensitive to environmental demands that deviated from optimal conditions than were redundant fleets.

**Validation II**: The next set of validation tests utilize similar conditions as before except that non-decomposable variation is now permitted in the set of validation scenarios. More precisely, differences between scenarios are decomposable when comparing scenarios within the training set, but not decomposable for comparisons within the validation set or for comparisons across sets. This is a harder validation test to satisfy, firstly because the validation scenarios do not meet the assumptions implicit in the training data. Moreover, decomposition requirements restrict scenarios to compact regions of scenario space and in removing this restriction, the scenarios can now be more distinct from one another, even if their distance from the seed position remains unchanged. Under these validation conditions, we see that the robustness of redundant fleets is considerably degraded. Degenerate fleets also suffer degradation in robustness, however they also display substantially better responses towards this type of novelty than do redundant fleets.

### Other experimental conditions
Our findings vary somewhat depending on the parameter settings that are used to generate the validation tests. For both validation sets, as the distance between the training seed and validation seed grows, robustness decays in both types of fleets although this is more rapid for the redundant fleets. As the volatility is reduced in the training set, the robustness advantage of degenerate systems also decreases; in the limit where there is only a single member in the set of training scenarios, both fleet classes always evolve to optimal robustness.

## Evolution in Environments with Non-Decomposable Volatility
In strategic planning, it is unlikely that anticipated future states (e.g. training scenarios) or the actual future states (e.g. validation scenarios) will involve compartmentalized variations that can be addressed separately. In the context of our model, this means that task requirements are more likely to change such that the covariance in task frequencies cannot be decomposed into a set of isolated relationships or sub-problems. The second set of validation scenarios evaluated in Figure 2b were formulated to capture this type of "non-decomposable uncertainty" as they allow correlations to arise between the frequencies of any pair of task types.

For these reasons, we focus our attention on the robustness and design adaptation characteristics for fleets that *evolve* under Validation II type non-decomposable environments. By tracking fleet robustness during the course of evolution, we now see in Figure 3a that degenerate fleets are becoming substantially more robust and do so more quickly than fleets evolved using reductionist design principles.

In our view, there are two factors that primarily contribute to these observed differences: i) evolvability of the fleet architecture and ii) differences in the robustness potential of the system. Conceptually, evolvability is about discovering design improvements. In Figure 3d, we record the probability distribution for the time it is taking the optimization algorithm to find new improvements to the fleet design. As can be seen, degenerate architectures are allowing for a more rapid discovery of design improvements than redundant architectures. A further analysis of fleet evolution confirms that design adaptation rates predominantly account for this divergence in robustness values. In Figure 3b we track degeneracy levels within the fleet and find that, when permitted, degeneracy readily emerges in the fleet architecture. We will show that the growth of degeneracy during evolution is a key factor in both the robustness and design adaptation of these fleets.

## Adaptation to Novel Conditions
One of the claimed advantages of degeneracy is that it supports a system in dealing with novelty; not only in system design but also in environmental conditions. To investigate this claim, we expose the fleets to an entirely new set of scenarios and allow the fleet designs to re-evolve (Figure 3a, generations 3000-6000). When high levels of degeneracy have already evolved within the fleet architecture, we see that the speed of design adaptation increases. When only redundancy is permitted, design adaptation after shock exposure at generation 3000 appears marginally worse. In our analysis of these results, we determined that the degraded adaptation rates in redundant fleet (re)evolution is an artifact of the optimization procedure and can readily be explained based on the following considerations: i) we use a population-based optimization algorithm, i.e. many fleet designs are being modified and tested concurrently, ii) the population of fleets is converging over time, i.e. differences between fleets or *diversity* get lost (Figure 3c), and iii) it is generally true, in both biology and population-based optimization, that a loss of diversity negatively affects a population's adaptive response to abrupt environmental change. We can confirm the population diversity effect by initializing populations at low diversity or more simply by reducing the population size, which results in redundant fleet robustness profiles that are indistinguishable when comparing evolution in old (generations 0-3000) and new (generations 3000-6000) environments.

In contrast, degenerate fleets evolve improved designs more rapidly in the new environment (Figure 3a) and this effect is not eliminated by altering the population size. However, when we restrict the total amount of degeneracy that is allowed to arise in the fleet (Figure 3b), the fitness profiles before and after shock exposure become increasingly similar (Figure 3a). Together, these findings support the hypothesis that degeneracy improves fleet design evolvability under novel conditions.

### Innate and Evolved Differences in the Robustness Potential of Degenerate Fleets

In our introductory remarks, we proposed that degeneracy may allow for pervasive flexibility in the organization of system resources and thereby afford general advantages for a system's robustness. To evaluate this proposal, in Figure 3d we introduce fleets that are not evolved but instead are artificially constructed (see Section 3) so that they have levels of degeneracy and redundancy that are similar to their evolved counterparts from the experiments in Figure 3a. Figure 4 reports the robustness of these (un-evolved) fleets against randomly generated scenarios. We see that innate advantages in robustness are associated with degenerate fleet architectures. We contend that this finding is not intuitively expected based on a reductionist view of vehicle capabilities given that: both degenerate and redundant fleets have the same number of vehicles (and the fleet design spaces are constrained to identical sizes), fleets are presented with the same scenarios, and vehicles have access to the same task capabilities.

It is interesting to ask whether this innate robustness advantage can explain entirely our previous results or whether the manner in which degeneracy evolves and becomes integrated within the fleet architecture is also important, as was proposed in the networked buffering hypothesis [24]. To explore this question, in Figure 4 we show robustness results of the final evolved fleets from Figure 3a but now robustness is reevaluated against randomly generated scenarios. Here we see that, when degeneracy evolves, a fleet's robustness to novel conditions becomes further enhanced, indicating that the organizational properties of degeneracy play a significant role in a system's robustness and capacity to adapt. Importantly, it suggests that the flexibility is tied to the history of the system's evolution yet remains pervasive can emerge without foresight, thus satisfying our third stated requirement for supporting innovation.  In contrast, the capacity to deal with novel conditions is unaffected by evolution in redundant fleets and remains generally poor for both evolved and randomly constructed fleet designs.

In summary, we have found that degenerate fleet designs can be broadly robust as well as highly adaptable; enabling these fleets to exploit new design options and to respond more effectively to novel environments.  Only when environments are stable and characterized to favor reductionist principles did we find redundant architectures to slightly outperform degenerate architectures. While these preliminary findings are promising, there are additional factors that should be considered when weighing the pros and cons of degeneracy design principles. In the remaining experiments, we begin to investigate some of the issues that arise within the context of our case study.

### Costs in the Design and Redesign of Fleets
Depending on how the planning problem is defined, it may be of interest to know how the fleet composition changes in response to a new set of environmental demands.  For instance, if we considered the initial conditions in our experiments to represent a pre-existing fleet, then it would be important to know the number of vehicles that are being replaced, as this would clearly influence the investment costs.

In this study, we treat the entire fleet being modeled as reflecting a new investment decision and thus do not consider changes in the fleet during evolution as having an influence on investment costs. To address fleet replacement issues would at the very least require the specification of vehicle replacement costs, new vehicle design costs, and the inclusion of an aggregate cost function as a second objective in the optimization problem, i.e. maximizing robustness while minimizing costs. While we do not consider here a multi-objective formulation of our planning problem, it is interesting to evaluate the relationship between the magnitude of fleet change and fleet performance within our previous experiments. In both proposing and exploring the theoretical merits of degeneracy, we have emphasized its role in facilitating the discovery of incremental design changes. We have found that degenerate fleets are highly evolvable and we further speculate that more adaptive options exist at each stage of evolution [49]. Thus we would expect that, without cost objectives in place, degenerate fleets would diverge rapidly from their original conditions. Interestingly, we find in Figure 5 that divergence is not substantially greater in comparison to reductionist design conditions and that more favourable cost-benefit trade-offs appear to exist for evolution under non-decomposable environments, *cēterīs paribus*.

For example, in decomposable environments an optimally redesigned redundant fleet was typically found by replacing 15% of the vehicles while degenerate fleets achieved near optimal performance when approximately 20% of the fleet is redesigned (Figure 5a). In complex environments, no fleet typically is able to discover an optimal redesign, however, the increased propensity to discover adaptive design options is shown to confer degenerate fleets with a performance advantage that becomes more pronounced as larger investments are considered (Figure 5b).

However, such conclusions do not factor in additional costs from the development of new vehicle designs or reduced costs that may come from economies of scale during the manufacturing and purchase of redundant vehicles. While such costs depend on the planning context, these factors will influence the advantages and disadvantages from degeneracy and should be explored further. While having noted the potential costs from degeneracy, one should not immediately conclude that fleet scaling places degeneracy design principles at a disadvantage. For instance, in Figure 5 we consider fleet evolution (under non-decomposable environments) at different sizes of our model. Here we see that the robustness advantage from degeneracy has a non-linear dependence on fleet size. Moreover, because degeneracy is a relational property, modifying the design of a small number of vehicles can have disproportionate influence on degeneracy levels within a fleet. For instance, in one of the data sets in Figure 3a only 20% of the fleet is permitted to deviate from its initially redundant architecture yet we observe considerable improvements in degeneracy and fleet robustness. In short, the careful redesign of a small subset of the fleet could provide considerable advantages, particularly for large fleet sizes.

## Multi-Scaling Effects

Our study has focused its attention on a dynamic optimization problem that arises at the strategic level of fleet planning, i.e. involves vehicle investment decisions over many years. However, as is also true in biological models, strategic planning constitutes only one facet of a rich multi-scaled topic that is likely to have important cross-scale interactions [72] [36] [73].

For instance, in our estimation of fleet performance (fitness evaluation), it was necessary to determine how vehicle resources can be matched to the task requirements for each scenario. While decision support tools might view this as a simulation [65], in operations research this simulation would be formulated as a resource assignment problem. Regardless of how the fitness evaluation is viewed conceptually, architectural properties of the fleet are likely to have non-trivial consequences in the assignment of vehicles to tasks. We have already demonstrated that the fleet architecture (as well as the environment) influences the quality of task assignment solutions that can be evolved and we suspect that it will also influence the difficulty associated with implementing these solutions, i.e. matching tasks with suitable vehicles.

In our experiments, task assignment is approximated using a local decision-making heuristic (formally described in Section 3). This heuristic is used in order to simplify the experiments while maintaining a rough analogy to how these decisions are sometimes made, i.e. by vehicle owners (management) instead of optimization algorithms. The heuristic includes an "owner" of each vehicle that distributes a vehicle's resources (time) over suitable tasks. However, as other owners make their own decisions, the relative importance of remaining unfulfilled tasks can change and owners may decide to readjust their vehicle's tasks, thus (indirectly) adapting to the decisions of others. In the following, we investigate how a fleet's architecture influences the amount of time required to complete this task assignment procedure and similarly, how placing restrictions on the time allotted to the procedure, i.e. changing the *maximum simulation runtime*, influences the performance results for different fleet architectures.

In Figure 6a, we evolve fleets using different settings of the maximum simulation runtime and record how this alters the final robustness achieved during evolution. In the limiting case where vehicle task assignments are never updated, a fleet has no capacity to respond to changes in the environment and performance is entirely genetically determined, i.e. vehicles are not functionally plastic. In this case the problem collapses to one that is equivalent to optimizing for a single scenario (i.e. the mean task requirements of the scenario set) and the two types of fleets evolve to the same poor performance. When fitness evaluation is extended to allow short simulation times (few decision adjustments), degenerate architectures display modestly better matches between vehicles and tasks and this continues to improve as simulation time is increased and robustness converges to near optimal values (Figure 6a).

Earlier experiments in this study did not impose restrictions on the simulation runtime, i.e. task assignment heuristics were run until no decisions could be readjusted. Here we see that simulation constraints may have influenced our results if decision adjustments had been sufficiently reduced, e.g. to less than 20. To evaluate the task assignment efforts that took place in our earlier experiments, Figure 6c plots the frequency distribution for the number of decision adjustments that occur as a fleet responds to a scenario. Here we see that the discovery of a stable assignment can take considerably longer for degenerate fleets than for redundant fleets.

However, we also have seen in Figure 6a that constraining the assignment runtime had only modest repercussions to performance when the fleets are forced to evolve under these constraints. We have found two factors that contribute to these seemingly contradictory findings. First, the largest fitness improvements that result from decision adjustments predominantly occur during the early stages of a simulation. This is shown in Figure 6b where fitness changes are recorded as decision adjustments are made. While some fleet-scenario combinations potentially require many decision adjustments, the fleet still provides decent performance if the simulation is terminated early. We also speculate that a second contributing factor could be that evolution under restricted runtimes influences how the architecture evolves. With simulation runtime restricted, evolution may prefer fleet architectures that find decent assignments more quickly.

To investigate this conjecture, we evolve fleets with simulation runtime restricted to at most ten decision adjustments. We then remove this restriction and in Figure 6d record the total number of decision adjustments as the fleets respond (without runtime restriction) to new scenarios. Compared to the results in Figure 6c, we see a large reduction in runtime distributions indicating that fleets have been designed to be more efficient in task assignment. Figure 6b also indicates that higher quality vehicle assignments are found more quickly for degenerate fleets optimized under these constrained conditions.

In summary, an exploration of the underlying resource assignment problem indicates that important interactions exist in fleet performance characteristics as they are observed at different timescales of our planning problem. However, in the context of the distributed decision-making heuristics that were implemented, we have generally found that degeneracy principles can result in high robustness without incurring large implementation costs. Importantly however, satisfactory cross-scale relationships depend on whether short-timescale objectives (i.e. simulation runtime) are being accounted for during fleet planning.

# 5. Discussion

### Taking Inspiration from Biology
Fleet degeneracy transforms a trivial resource allocation problem that, in principle, can be solved analytically into a problem with parameter interdependencies that is no longer analytically tractable for large systems. From a traditional operations research perspective, degeneracy should thus be avoided because it increases the difficulty of finding a globally optimal assignment of resources. However, despite increased complexity in centralized resource control, we have found that simple distributed decision-making heuristics can provide excellent resource allocation assignments in degenerate fleets. While optimality is not guaranteed, the superior fleet design easily lends itself to a more effective allocation of resources compared with the globally best solutions for fleets designed based on reductionist principles.

By exploring this conflict between design and operational objectives, our study has identified conditions that support but also limit the benefits that arise from degeneracy. These conditions appear to be in agreement with observations of distributed control in biological systems where trait robustness generally emerges from many localized and interrelated decisions [27]. How these local actions translate into system traits is not easily understood by breaking down the system into its respective parts. This lack of centralized control thus has put into question whether biological principles are compatible with engineering and socio-technical systems where centrally defined global objectives play an important role in assessing and integrating actions across an organization. However, and in contrast to common intuition, we have found some evidence to suggest that difficulties in centralized control do not preclude the possibility of coherent system behaviours which are locally determined yet also driven by globally defined objectives.

Reductionist design principles should be well-suited for addressing planning problems that can be readily separated into sub-problems and where variations in problem conditions are bounded and predictable. On the other hand, we propose that biological design principles such as degeneracy should be better suited for environments that are not easily decomposed and that are characterized by modest levels of unpredictable novelty. However, when environmental changes become too extensive or diverge too rapidly from historical conditions (relative to the pace of design adaptation), we suspect that no class of systems will be able to meet the basic performance requirements set out for a fleet.

Furthermore, with fitness proving to be a stronger selection criterion, this helps to dispel potential criticisms that degeneracy only emerges under weak selection, thereby having some added relevance to the conclusions drawn from our findings, as we discuss later.

## Complex Systems Engineering

As a design principle, degeneracy could be applicable to several domains, in particular in circumstances where

1) distributed decision-making is achievable (e.g. in markets, open-source software, some web 2.0 technologies, human organizations and teams with a flat management structure);

2) individual motivations can be aligned with system objectives; and,

3) agents are agile (i.e. quickly responsive to their environment) and functionally versatile.

When these conditions are met, designing elements with partially overlapping capabilities can dramatically enhance system resilience to new and unanticipated conditions as we exemplified in our study on field vehicle fleets. On the other hand, if there is a need to maintain centralized decision-making and system-level transparency, or in cases where historical bias favoring reductionism is deeply ingrained, implementing degeneracy design principles is likely to prove difficult and possibly unwise.

While multi-functionality is not a typical feature in systems made up of simple elements, it is commonly observed in more sophisticated machines, subsystems, and particularly military field vehicles[2] and furthermore is a requirement for the degeneracy design principles being explored in this study.

While there is a growing awareness of the role played by degeneracy in biological evolution, these insights have not yet been taken into account in the application of nature-inspired principles towards real-world problems. We propose that our findings can shed new light on a diverse range of research efforts that have taken insights from biology and complexity science to address uncertainties arising in systems engineering and planning [16] [17] [18] [19] [20] [21] [22] [23]. A full treatment of the relationship between degeneracy and other concepts developed in complexity science is outside the scope of this article. However, we stress that what we have proposed here does not constitute an isolated set of principles. Instead, degeneracy is a system property that:

1) based on both empirical evidence and theoretical arguments, can facilitate the realization of other important properties in complex (evolved or planned) systems, including weak/loose coupling [74] [75], multi-scaled complexity [3] [76], resilience through far from equilibrium dynamics [36], robustness within outcome spaces [77], the establishment of "solution rich configuration spaces",  and evolvability [3] [45] [74];

---

[2] The ability to carry out different task types is a common feature of land vehicles. Different tasks may involve different terrains, road and weather conditions, driver certifications, payloads, security conditions, as well as different types of activities such as transport, convoy security, patrol, communication, repair, medical services, etc.

2) depends on the presence of other basic system properties that were implemented in our experiments such as distributed decision-making [77], feedback [62] [78], modularity/encapsulation [74], protocols/interfaces/tagging [7] [62], exploration [45], agent agility  [21] [78], and goal alignment or coordination within a multi-scaled system [3] [78].

## Dynamic Capabilities of a Firm

One notable example where degeneracy could be particularly relevant is human organizations. This could be especially true in situations where fast and effective responses to unexpected conditions in a market or regulatory environment, to disruptive technologies, or to military threats is important to group survival. In organization science, such adaptability is discussed as *dynamic capabilities* (cf. [79] [80] [81]). In [81], the authors define dynamic capabilities as "the abilities to reconfigure a firm's resources and routines in the manner envisioned and deemed appropriate by its principal decision-maker(s*)*." The guiding principles and conditions that support dynamic capabilities have been an area of active research in the last decade, at least in part due to the growing volatility of the environments in which many businesses operate [82] [83] [84] [85] [86] [87].  As a consequence of these business trends, researchers have begun using principles from biology [88] [89] [90] [91] and complexity science [92] [93] [94] [95] to understand and improve upon the dynamic capabilities of a firm.

Given the evidence that suitable parallels exist, e.g. in the exceptional agility and versatility of humans and the flattening of management structures in many firms, a principled approach to enhancing degeneracy within an organization can have a considerable influence on an organization's adaptive capabilities. Thus we argue that it is important to improve our understanding of the conditions where multi-skilled workers, versatile problem-solvers, and so called "attention deficit" brains, can support organizational cohesion [75] [96], organizational integration (e.g. through the formation of dynamic assemblies/modules/teams [97]; see below), and flexibility in the construction/execution of organizational capabilities.

Within a team environment, any improvements in flexibility will only be realized however if consideration is also given to other practical realities such as the significance of relationship maintenance and trust in social/professional interactions. To rely on the strengths of others or to support others in their tasks requires a mutual understanding of intention, motivation, and commitment. For individuals with highly versatile skill sets, their ability to establish and maintain personal relationships and communication channels will determine their ability to utilize these diverse strengths to the best effect in the team and organization.  This also implies that teams cannot exist as compact modules with little interaction with their outside world, only becoming flexible in times of crisis. Instead, adaptive teams must exist as fluid entities that are in the habit of assembling modified team compositions and structures that are constructed partly based on the evolving needs of the team and partly based on the established relationships of versatile and degenerate team members across the broader organization. The benefits from these ideas should become most relevant to highly dynamic and unpredictable environments, for instance in the Australian Defence Force's *combined arms teams* where adaptation, trust, and timing within novel combat situations can determine mission success.

## Reengineering in Complex Systems

Our introductory remarks regarding system's engineering intentionally emphasized distinctions between designed and evolved systems, however engineering in practice can take on features of both. Engineering is not simply drawing up a plan and implementing it. Human-constructed systems typically coevolve with their environment during planning, execution, and even decommissioning activities. While biological terminology is not common to this domain, several attributes related to degeneracy are found in socio-technical systems and are sometimes intentionally developed during planning.

For instance, functional redundancy is a common feature in commercial/social/internal services when a sustained function under variable conditions is critical to system performance or safety, e.g. communications and flight control for commercial aircraft. Functional redundancy is rationally justified in these cases based upon its relationship to service reliability (see Section 1) which is analogous to the principles underpinning economic portfolio theory, i.e. risks are mostly independent while returns are additive. However if nothing else, this study and other recent studies have shown that a significant amount of the robustness derived from degeneracy has origins that extend beyond simple diversity or portfolio effects and is conferred instead as an emergent system property. For instance, previous work has theorized and found evidence that the enhanced robustness effects from degeneracy originate from distributed compensatory pathways or so called *networked buffering* [24].

Moreover, it is not simply reliability in function but also the relationship between robustness and innovation that makes degeneracy especially important to systems engineering. To better understand this relationship between degeneracy, robustness, and innovation, we have taken an interdisciplinary approach that combines system's thinking, biological understanding and experimentation using agent-based simulations. In ongoing research, we are using this approach to explore how degeneracy might facilitate successful organizational change within a firm as well as successful reengineering for systems that display strong functional interdependencies in the mapping between agent actions and system capabilities. Below we touch upon some of these issues and discuss why we believe degeneracy could lead to new paradigms both for reengineering and the dynamic capabilities of a firm.

**Complexity in Engineering**
While the term *complexity* generally relates to the interdependence of component behaviour/actions/functions, it is an otherwise ambiguous term and there is no consensus as to its meaning or measurement, e.g. [70] [98] [99] [100] [101] [102] [103]. In engineering, complexity is often used to describe sophisticated services involving several interdependent actions that are performed by single-purpose devices in specific ways, places, and times. This form of complexity, sometimes referred to as *specification complexity*, increases a system's sensitivity to novel conditions in a straightforward and well-understood manner.

Starting with a single device, the number and exactness of functional requirements/specifications/constraints generally influences the proportion of operating conditions that will meet these requirements. While the tradeoff between operating constraints and operational feasibility is not always simple (e.g. linear, monotonic) even for a single device, it is widely acknowledged that multi-component services tend to become more sensitive to novel internal and external conditions as more components are added to the system that co-specify the feasible operating conditions of other system components. In particular, the operating requirements placed on each component become more exacting as its function becomes more reliant on the actions/states/behaviours of others, e.g. through direct interaction, through sharing or modifying the same local resources, or indirectly through failure propagation. Complex services then tend to become more fragile to atypical component behaviours or atypical events because a greater proportion of events can exceed the operational tolerance thresholds in one or more devices, with the propagation characteristics of these threshold-crossing events determining the likelihood of sub-system and system-wide failure. To reduce the frequency of such failures, a design approach is generally taken that assumes predictability and relies on carefully placed backup devices and other fail-safe principles.

These common operational demands have an important impact on the challenges faced during redesign/reengineering. When systems are designed from single purpose devices that are each uniquely suitable for a system-critical function, this establishes a tight coupling between system performance, the reliability of a function, the continued normal operation of the device providing that function, and the continued compatibility of that device with other interacting devices.[3] Novel redesign of devices is thus constrained by a need to properly interact/communicate with other specific devices. With engineering driven to maximize efficiency and performance, small design adjustments are repeatedly made over time to improve efficiency under standard operating conditions, i.e. the system's overall design becomes well-adapted to its environment. As a system matures (evolves) in this way, the available degrees of freedom for further modifying a system's design naturally become limited.

This stagnation in redesign will lead to tension if the environment and system priorities change while options for design modification remain limited. Eventually a failure to meet system-level goals makes reengineering projects unavoidable, however redesign constraints remain and thereby force design modifications to be required in many components simultaneously; a point well illustrated in the dramatic redesign of complex software, e.g operating systems. A large reengineering effort often runs roughshod over the accumulated and highly contextual knowledge that was built during maturation, causing large reengineering and change management projects to often appear as failures when compared to prior system performance.

---
[3] While functional redundancy is sometimes designed into a system, it is almost always treated as a backup device that is not utilized under standard operating procedures.

**Complexity in Biology**

Highly sophisticated services also arise in biological systems and require many different sub-functions and process pathways to be executed. However, the building blocks of biological systems are not single purpose devices with predefined functionality and instead display multi-functionality, functional plasticity, and degeneracy.

While occasional slowdowns in the tempo of adaptation is inevitable and occurs in biological evolution as well (e.g. under stabilizing selection), there is currently no evidence that biological systems display the same built-up tension or sensitivity to incremental design changes, and we believe this is because degeneracy affords a weaker coupling between the functions performed and the components involved in achieving them [45]. Within an abstract design space or fitness landscape, one might say that engineered systems find themselves on isolated adaptive peaks while biological systems reside on connected neutral plateaus. While many complexity science researchers have used this rugged fitness landscape metaphor to advocate the need for more disruptive and explorative design changes, this is neither required nor observed in biological evolution.

In biological evolution, continued species survival requires that incremental adaptive design changes can be discovered that do not lead to a propagation of redesign requirements in other components in the system, i.e. macro-mutation is a negligible contributor to the evolution of complex species. Instead, single heritable (design) changes are found that lead to (possibly context-specific) novel interaction opportunities for a component, flexible reorganization of component interactions (that still maintain core functionalities), and in some cases a subsequent compounding of novel opportunities within the system [28].[4] In other words, the requirement is one of incremental changes in design and compartmentalized, but not necessarily incremental, changes in system behaviour. As elaborated upon in Section 1, we believe that many of the conditions needed to reconcile these competing requirements for robustness and adaptability in biology depend on the existence of degeneracy. On the other hand, the adaptability of biological systems is probably not derived solely from inherent structural properties either. Being in the habit of actively utilizing plastic responses and combining system parts in different ways also appears to be important. This conjecture is supported by evidence that successful innovations in biology often rely on past and present routines of utilized flexibility [45].

Within a (meta) stable environment, these proposed limitations to classic engineering paradigms may appear intermittent and not of grave concern. Indeed, the limitations we have described are not expected to be a problem for complex engineered systems so long as the environments of the constituent elements are highly stable and redesign is not coupled to the system's long-term viability. However the underlying associated assumptions of system closure, equilibrium conditions, and stable environments are becoming less suitable as the pace of technological change increases [74] [82] [83] [84] [85] [86] [87, 104] [105] [106] and as product development times continue to shrink [107] [108] [109] [110] [111] [112] [113] . It is the growth of volatility, uncertainty, and non-equilibrium conditions within many industry sectors that makes the issues we have raised here of particular relevance to present day reengineering challenges.

## 6. Concluding Remarks

In this article, we have investigated the properties of degenerate buffering mechanisms that are prevalent in biological systems. In comparison to their engineering counterparts, these buffering mechanisms were found to afford robustness to a wider range (both in type and magnitude) of perturbations and do so more efficiently due to the manner in which these buffers interact and cooperate. While seemingly paradoxical, we also hypothesized how the same mechanisms that confer trait stability can also facilitate system adaptation (changes to traits) under novel conditions. With the design of transportation fleets taken as a case study, we reported evidence supporting this hypothesis, demonstrating that the evolution of degeneracy results in fundamentally different robustness and design adaptation properties within fleets.

In looking to tackle real-world problems using inspiration from biology, it is important to determine whether sufficient parallels exist between the problem and the biological system of interest. Here we have proposed that the investment decisions and subsequent operation of some complex engineered systems consisting of versatile semi-autonomous agents provide a general domain where these requisite conditions are met and where degeneracy design principles could prove advantageous. There are a number of systems that can be characterized in this manner, with strategic planning for field vehicle fleets provided as one illustrative example.

---

[4] Some examples of robust yet flexible behavior that is believed to enhance biological evolvability are provided by Kirschner and Gerhart in [45].

# Acknowledgements

This work was partially supported by a DSTO grant on "Fleet Designs for Robustness and Adaptiveness" and an EPSRC grant (No. EP/E058884/1) on "Evolutionary Algorithms for Dynamic Optimisation Problems: Design, Analysis and Applications."

# Figures

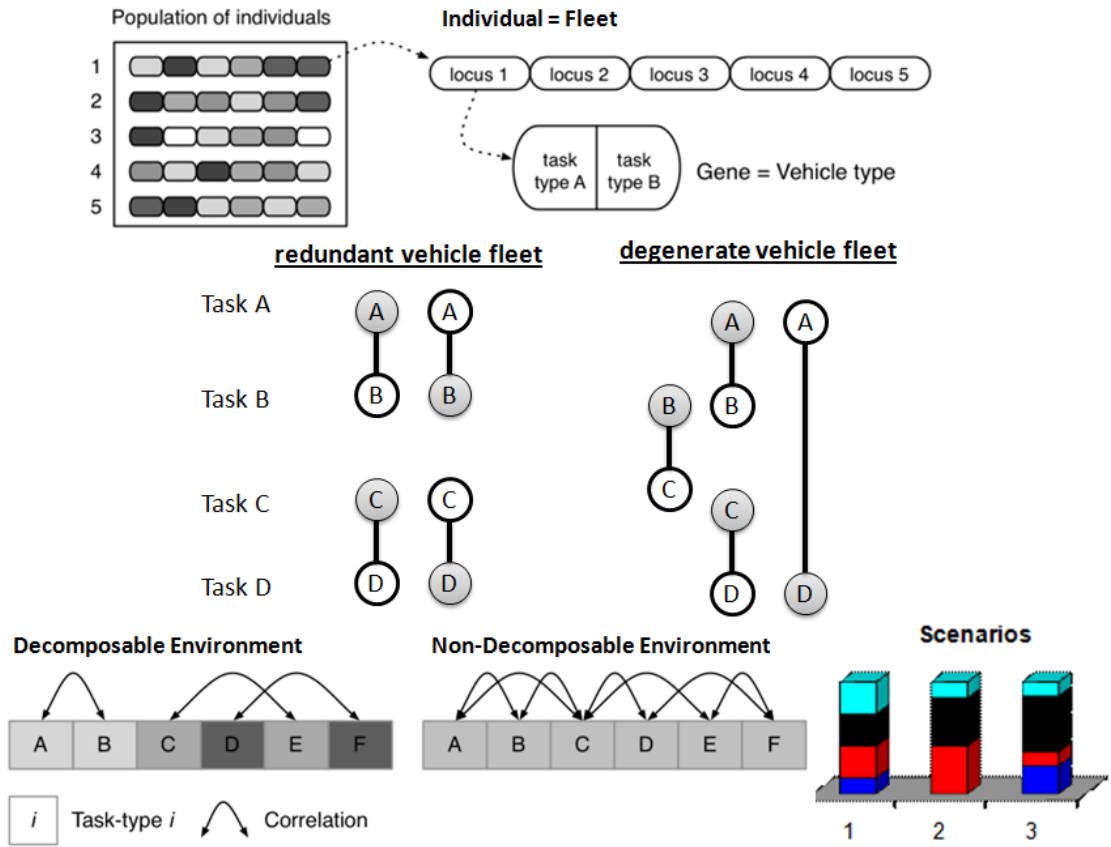

Figure 1: Top panel) Fleet encoding in genetic algorithm. The genetic algorithm evolves a population of N individuals, each of which represents a complete fleet of vehicles: at each locus, an individual/fleet has a specific vehicle type as specified by two distinct task types. Middle panel) A simple example highlighting the differences between sets of redundant and degenerate vehicles: given four vehicles (each vehicle represented as a pair of connected nodes) and four task types – A, B, C, D –, the fleet on the left-hand side consists of two sets of identical vehicles. The fleet on the right-hand side consists of four unique vehicles with partial overlap between their functionalities (i.e., task types); in both cases, each task is represented to the same degree. Bottom Left Panel) An illustration of task-type frequency correlations that characterize the variation in decomposable and non-decomposable environments. Bottom Right Panel) Environmental scenarios differ from one another in the frequency of task types (differentiated by color) but not the total number of tasks (vertical axis).

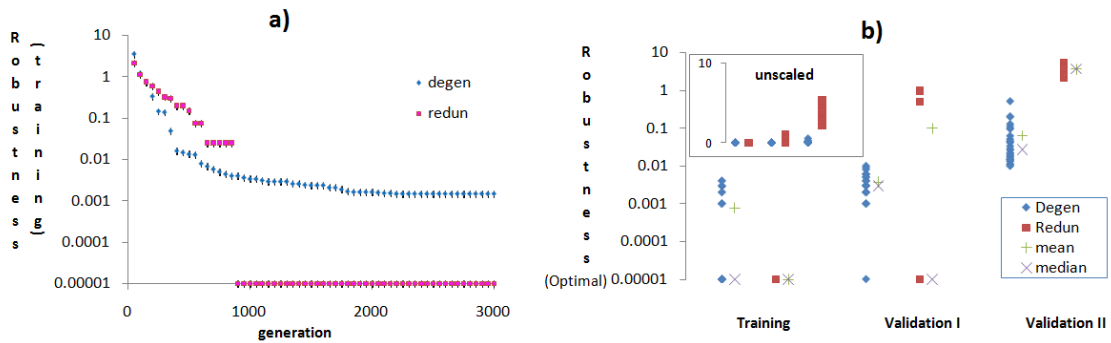

Figure 2 Panel a: Robustness profiles of degenerate and redundant fleet designs evolved in environments with decomposable variation (*training scenarios*). Panel b: Robustness is then reevaluated within new *validation scenarios*. "Training" shows robustness values of all 30 fleets at the end of evolution against the training scenarios of Panel a. "Validation I" and "Validation II" show robustness values for the 30 evolved fleets against two different types of validation scenarios (described in main text). The insert shows the same results as the main panel but on a linear (not a logarithmic) robustness scale.

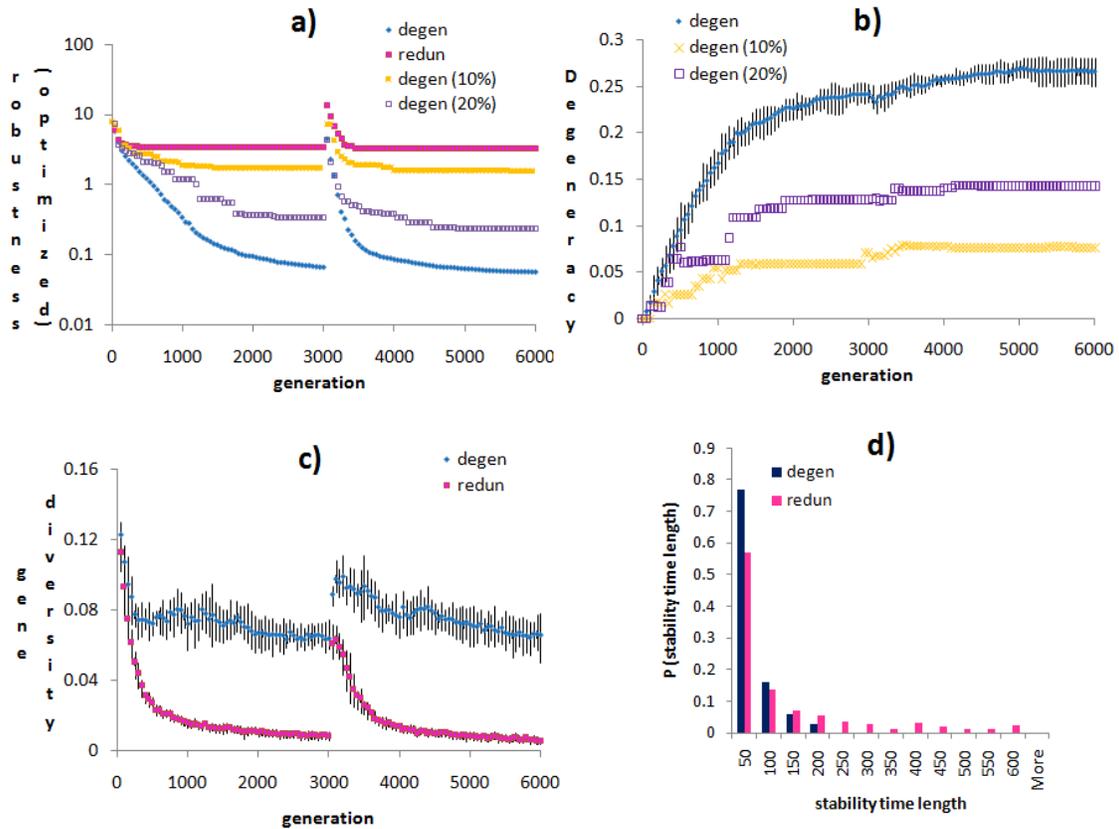

Figure 3 Panel a: Robustness profiles of degenerate and redundant fleet designs evolved in environments with non-decomposable variation and exposed to a shock (i.e. an entirely new set of scenarios) at generation 3000. Characteristic results are also shown for experiments where only a predefined percentage of vehicles can be designed to no longer conform to the initial redundant fleet architecture and thus allow for some restricted level of partially overlapping capabilities within the fleet. Panel b: Degeneracy is measured (for fleets where it can emerge) during the course of evolution. Panel c: Gene diversity during the evolution of Panel a; measured as the proportion of vehicles that have changed in pair-wise comparisons of fleets within the population. Panel d: histogram for the number of offspring sampled before an improvement is found (time length). Sampling is restricted to the evolution of fleets throughout the first 3000 generations of Panel a.

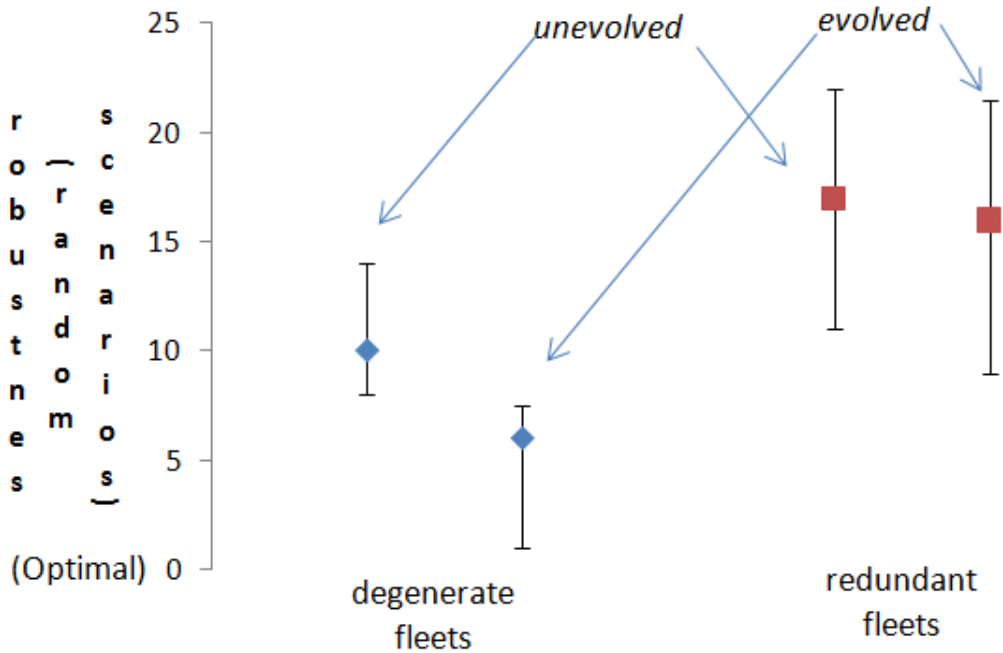

**Figure 4** Comparisons of the robustness of evolved and un-evolved fleets towards randomly generated scenarios. Un-evolved degenerate fleets were constructed with levels of degeneracy as large as evolved fleets.

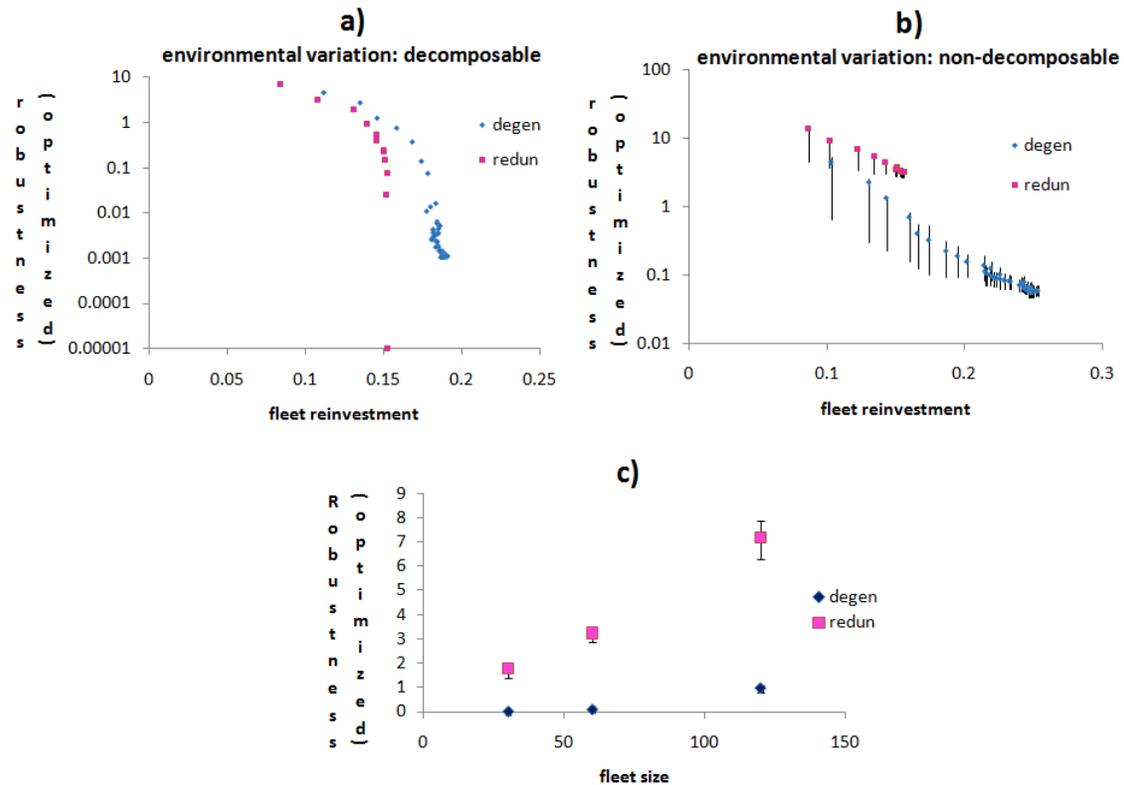

**Figure 5 Panel a:** Robustness of evolved fleets plotted against the proportion of vehicles that have changed when comparing an evolved fleet with its original design (at gen=0). Evolution takes place within a decomposable environment. **Panel b:** Same as Panel a but with evolution taking place in non-decomposable environments. **Panel c:** Comparisons of the evolved fleet robustness for degenerate and redundant architectures at different fleet sizes. In these experiments the fleet size, the number of task types $T$, random walk size, and maximum generations are all increased by the same proportion.

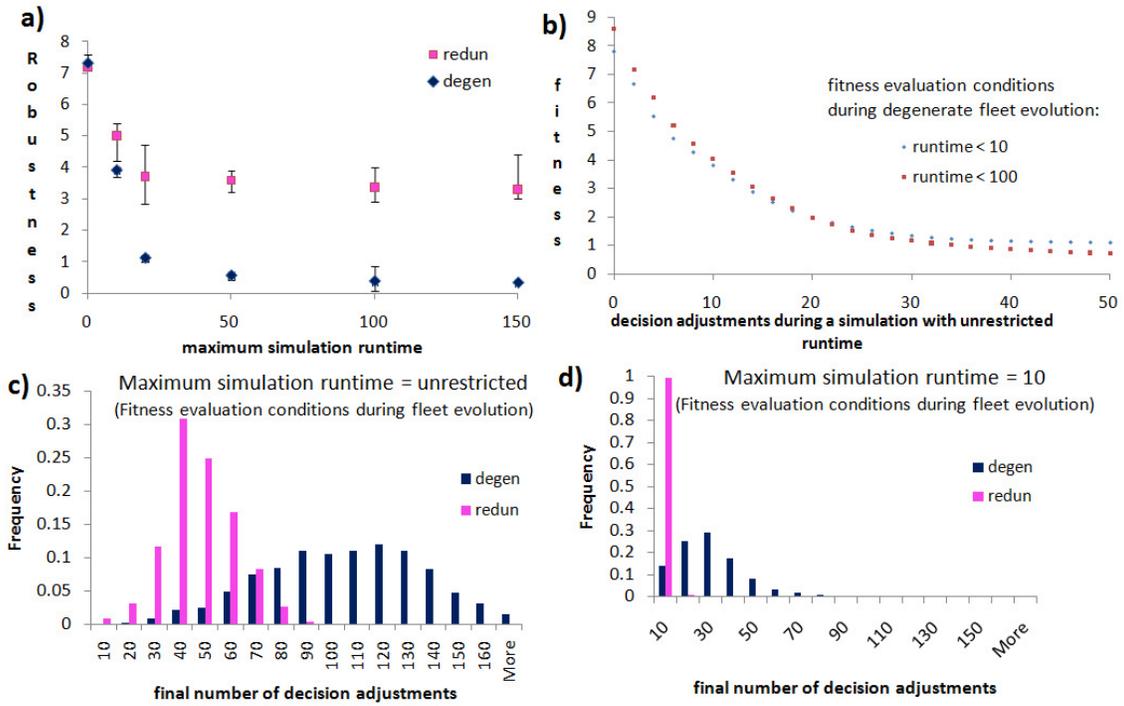

**Figure 6 Panel a:** Robustness of fleets that have been evolved with different restrictions on maximum simulation runtime. **Panel b:** Fitness of a fleet is evaluated in a single scenario where fitness is recorded during a simulation, i.e. as vehicle decision adjustments are made. The results are shown for degenerate fleets that have evolved in conditions where the maximum simulation runtime is 100 and 10 readjustments. **Panel c:** Actual runtime distribution for fleets evolved under unrestricted runtime conditions **Panel d:** Actual runtime distribution for fleets evolved under a maximum simulation runtime of 10, but where the distribution is being evaluated with these restrictions removed.